\documentclass[preprint,preprintnumbers,amsmath,amssymb]{revtex4}
\usepackage{graphicx}

\begin{document}

\title{Interpretation of Hund's multiplicity rule for the carbon atom}

\author{Kenta Hongo$^1$, Ryo Maezono$^2$, Yoshiyuki Kawazoe$^1$, and
Hiroshi Yasuhara$^1$}

\affiliation{$^1$ Institute for Materials Research, Tohoku University,
Sendai 980-8577, Japan}

\affiliation{$^2$ National Institute for Materials Science, Sengen
1-2-1, Tsukuba 305-0047, Japan}

\author{M. D. Towler, and R. J. Needs}

\affiliation{TCM Group, Cavendish Laboratory, University of Cambridge,
Madingley Road, Cambridge CB3 OHE, United Kingdom}

\date{\today}

\vspace{0.5cm}

\begin{abstract}
Hund's multiplicity rule is investigated for the carbon atom using
quantum Monte Carlo methods.  Our calculations give an accurate
account of electronic correlation and obey the virial theorem to high
accuracy.  This allows us to obtain accurate values for each of the
energy terms and therefore to give a convincing explanation of the
mechanism by which Hund's rule operates in carbon.  We find that the
energy gain in the triplet with respect to the singlet state is due to
the greater electron-nucleus attraction in the higher spin state, in
accordance with Hartree-Fock calculations and studies including
correlation.
The method used here can easily be extended to heavier atoms.

\end{abstract}

\maketitle

\section{Introduction}
\label{section:introduction}
\renewcommand{\thefootnote}{\fnsymbol{footnote}}

From an analysis of atomic spectra, Hund found that the electronic
configuration of the lowest energy is the one with the highest spin
multiplicity $S$.  Hund's multiplicity rule \cite{Hund} explains
almost all the configurations of the ground states of atoms, ions, and
molecules and even their low-lying excited states in most cases.  Even
simple models such as frozen orbital Hartree-Fock (HF) theory
generally adhere to the rule, although the mechanism by which Hund's
rule is obeyed depends on the accuracy of the theoretical description.
A widely accepted mechanism given originally by Slater \cite{Slater}
attributes it to a reduction in the electron-electron repulsion energy
$V_{\rm {ee}}$ in the highest $S$ state.  This mechanism is in fact
incorrect as it is based on the incorrect assumption of choosing the
same orbitals for states with different $S$ (frozen orbitals).  This
assumption leaves the kinetic energy $T$ unchanged for different spin
states, which cannot be correct as each eigenstate must satisfy the
virial theorem.  The virial theorem states that $2T+V=0$, where $V$ is
the total potential energy, which shows that if the potential energy
is decreased the kinetic energy must increase.  If the orbitals are
allowed to relax self-consistently they are found to depend on $S$
states.\cite{Fraga,Davidson_1965,Katriel1,Katriel2,Colpa,Koga,Boyd1,Boyd2,Boyd3,Boyd4,Boyd5}
The correct explanation of Hund's multiplicity rule is in fact that
the higher spin state reduces the electron-nucleus interaction energy
$V_{\rm {en}}$.  Interestingly it turns out that states with larger
$S$ have higher values of $V_{\rm {ee}}$, in direct contradiction to
the Slater mechanism.

Davidson \cite{Davidson_1965} confirmed that Hund's multiplicity rule
derives from a reduction in the electron-nucleus energy within the
self-consistent HF approximation for the low-lying excited states of
the helium isoelectronic series. The self-consistent HF method has
been applied to explain Hund's multiplicity rule for all atoms (with
the exception of Zr ($Z$=40) for which HF theory predicts an
incorrect ground state)
\cite{Fraga,Pauncz,Katriel1,Katriel2,Colpa,Koga}, giving the same
stabilizing mechanism.  Boyd\cite{Boyd1,Boyd2} considered the same
systems as Davidson but within a variational approach. He explained
why larger values of $S$ give lower energies as follows.  Larger $S$
implies more parallel spin pairs and hence a larger exclusion holes
around each electron.  Consequently, electrons can avoid each other
more effectively and move closer to the nucleus, which reduces $V_{\rm
{en}}$.  We refer to this as the \textit{less screening} mechanism.
We should expect that when a mechanism lowers $V_{\rm {en}}$ it should
also increase $V_{\rm {ee}}$, because the electrons will on average be
closer to one another.  The less screening mechanism is therefore in
contradiction to Slater's argument which implies that $V_{\rm {ee}}$
will be lower in the triplet state.  Configuration interaction (CI)
calculations have shown that the same mechanism operates in the
low-lying excited states of small molecules such as ${\mathrm
{CH}}_2$, ${\mathrm H}_{2}{\mathrm {CO}}$, and the first-row hydrides,
LiH, BH, NH, and FH.\cite{Boyd3,Boyd4,Boyd5}

The origin of Hund's multiplicity rule for heavier atoms has mostly
been examined at the HF level.  It is interesting to study the origin
of Hund's multiplicity rule in heavier atoms beyond the HF level.
However, the conventional approaches such as CI are difficult to apply
for the following reasons: (1) The wave function must be constructed
carefully from the configuration state functions (CSFs) giving the
largest contributions to the correlation energy, which becomes a
cumbersome task for heavier atoms. (2) A huge number of CSFs are
required to give a sufficiently good description of the correlation,
particularly the electron-electron cusps.
In this paper we report a promising approach to this problem
applied to the carbon atom as a prototype using quantum Monte Carlo
(QMC) approaches such as the variational Monte Carlo (VMC) and more
accurate diffusion Monte Carlo (DMC) methods.

\section{Description of the calculations}
\label{section:describe_calculations}
%
In VMC, expectation values are calculated using an approximate
many-body wave function, the integrals being performed by a Monte
Carlo method.  The approximate wave function normally contains a
number of variable parameters, whose values are obtained by an
optimization procedure.

DMC is a stochastic projector method for solving the imaginary-time
many-body Schr\"{o}dinger equation.\cite{foulkes_2001} Exact
imaginary-time evolution would in principle give the exact
ground-state energy, provided the initial state has a non-zero overlap
with the exact ground state.  However, the stochastic evolution is not
exact and the symmetry of the final state would not be fermionic, but
bosonic.  To maintain the correct fermionic symmetry the fixed-node
approximation \cite{Anderson,Reynolds} is used in which the nodal
surface of the wave function is constrained to equal that of an
approximate guiding wave function.  If the nodal surface of the
guiding wave function is accurate, fixed-node DMC will give a very
accurate energy.  The fixed-node DMC energy is less than or equal to
the variational energy calculated with the guiding wave function and
greater than or equal to the exact energy.

We used a single-determinant Slater-Jastrow guiding wave
function.\cite{Jastrow} The single-particle orbitals were obtained
from HF calculations using the GAUSSIAN98 code \cite{Frisch} and a
6-311G++(3df,2pd) basis set.  At a nucleus the exact HF orbitals have
cusps such that the divergence in the potential energy is cancelled by
an equal and opposite divergence in the kinetic energy.  Orbitals
expanded in a Gaussians basis set cannot have cusps and the sum of the
contributions from the kinetic and electron-nucleus energies diverges
at the nucleus.  In practice one finds that this results in wild
oscillations in the local energy when an electron approaches the
nucleus, which can lead to numerical instabilities in DMC
calculations.  To solve this problem we make small corrections to the
single particle orbitals close to the nuclei which impose the correct
cusp behavior.\cite{ma_2004}

Electron correlation is introduced by multiplying the determinant by a
Jastrow function which depends explicitly on the inter-electronic
distances and obeys the electron-electron cusp conditions.\cite{Kato}
Our Jastrow factors take the form~\cite{williamson_1996}
\begin{equation}
\label{Eq.6}J\left( {\bf R} \right) = -\sum\limits_{i>j} 
\left[ \frac{A}{r_{ij}}\left[1-{\rm exp}\left(-\frac{r_{ij}}{F}\right)
\right] {\rm exp}\left(-\frac{r_{ij}^2}{L_0^2} \right) +
S_1\left({r_{ij}}\right)\right] - \sum\limits_{i,I}
S_2\left(r_{iI}\right) \ ,
\end{equation}
where the indices $i$ and $j$ denote electrons and $I$ denotes ions.
$F$ is chosen so that the cusp conditions~\cite{Kato} are obeyed,
i.e., $F_{\uparrow \uparrow} = \sqrt{2A}$ and $F_{\uparrow \downarrow}
= \sqrt{A}$, and $S_1$ and $S_2$ are cusp-less.  $S_1$ and $S_2$ are
expressed as polynomial expansions in the inter-particle distances.
We used a total of 12 variable parameters in the Jastrow factor whose
optimal values were obtained by minimizing the variance of the VMC
energy.\cite{Umrigar,Kent} The wave functions obtained by this
procedure were used as the guiding wave functions for our DMC
calculations.  We used a timestsep of 0.001 a.u. within the DMC
calculations, which gives very small timestep errors.  All the QMC
calculations were performed using the CASINO code.\cite{CAS01}

Arguments about the mechanism by which Hund's rule operates centre on
the values of the individual energy terms $T$, $V_{\rm {ee}}$ and
$V_{\rm {en}}$.  As illustrated by Slater's early treatment, results
which do not obey the virial theorem to high accuracy can give a
spurious interpretation of the rule.  It is therefore necessary to
calculate the energy terms using an accurate method, and in the case
of a stochastic method such as DMC, to evaluate them to a high
statistical precision.  Because there is no ``zero-variance property''
for the individual energy terms it is necessary to perform long Monte
Carlo runs, and in our calculations we have accumulated the results
over 200,000 steps.

The DMC method generates the mixed distribution, $\Psi_{v}\Phi_{0}$,
where $ \Psi_{v}$ and $\Phi_{0}$ are the guiding (VMC) and DMC wave
functions, respectively.  Evaluating the expectation value of the
Hamiltonian $\hat H$ (or any operator which commutes with it) with the
mixed distribution gives an unbiased estimate of the energy, i.e.,
\begin{equation}
\label{eqmixed}
\left\langle \hat {H} \right\rangle_m = \frac{\left\langle {\Psi_v}
\right|\hat H\left| {\Phi_0} \right\rangle}{\left\langle {\Psi_v}
| {\Phi_0} \right\rangle} = \frac{\left\langle {\Phi_0}
\right|\hat H\left| {\Phi_0} \right\rangle}{\left\langle {\Phi_0}
| {\Phi_0} \right\rangle} = \left\langle \hat {H} \right\rangle_p \;,
\end{equation}
where the subscript $m$ denotes the mixed estimator and $p$ denotes
the pure estimator.  However, for expectation values of operators
which do not commute with the Hamiltonian the mixed estimate is
different from, and generally less accurate than, the pure estimate.
An approximation to the pure estimate can be obtained by combining VMC
and DMC results in an extrapolated estimation.\cite{ceperley_1979}
The linear extrapolated estimator is
\begin{equation}
\label{eq3}
\langle \hat{O} \rangle_{l} = 2 \langle \hat{O} \rangle_{m} - \langle
\hat{O} \rangle_{v} = \langle \hat{O} \rangle_{p} + \mathcal{O} (
\Delta^2 ),
\end{equation}
where $ \Delta = \Phi_{0} - \Psi_{v}$.  
Other extrapolated estimators such as the squared extrapolation,
\begin{equation}
\label{eq4}
\langle \hat{O} \rangle_{s} = \frac{\langle \hat{O} \rangle_{m}^2}
{\langle \hat{O} \rangle_{v}} = \langle \hat{O} \rangle_{p} +
\mathcal{O} ( \Delta^2 ),
\end{equation}
can also be used, which have different error terms.  In this study we
found that the linear and squared extrapolations gave almost identical
results, and we will quote results only for the linear one.

\section{Results and Discussion}
\label{section:results}

In Table I we give VMC and DMC expectation values for the different
each energy terms and for the virial ratio for the lowest energy
triplet and singlet states of the carbon atom.  For comparison, the HF
and CI values of Pauncz \textit{et al.} \cite{Pauncz,Katriel2} are
also given. The exact non-relativistic energy of the ground state of
the carbon atom has been estimated to be -37.8450
a.u.\cite{davidson_1991} We can use this value and the Hartree-Fock
energy of -37.689 a.u. to compute the percentages of the correlation
energy recovered.  The CI calculations of Pauncz \textit{et al.}
\cite{Pauncz,Katriel2} recovered 45.6 \% of the correlation energy,
while a more modern CI calculation \cite{Noro} (which did not report
the individual energy terms) recovered 61.5 \%.  Our DMC calculations
recover 88.5 \% of the correlation energy.

For each of the VMC, DMC and extrapolated DMC calculations we obtain a
virial ratio within statistical error bars of its correct value of
$-V/T = 2$.  It is straightforward to show from the virial theorem
that the exact kinetic energy should be larger than the HF one in an
isolated system.  This inequality is obeyed by each of our
calculations but is disobeyed by the CI calculations of Lemberger and
Pauncz,\cite{Pauncz} whose calculations also show a significant error
in the virial ratio.  This indicates that the virial ratio provides a
sensitive test of whether the relationships between the different
energy components are correct.

Comparing the DMC and extrapolated DMC as results with the HF ones for
a particular state we find
\begin{eqnarray}
V_{\rm {ee}} & < & V_{\rm {ee}}^{\rm HF} \nonumber \\
V_{\rm {en}} & < & V_{\rm {en}}^{\rm HF} \nonumber \\
T & > & T^{\rm HF}.
\end{eqnarray}
The introduction of correlation reduces both the electron-electron and
electron-nuclear potential energies at the expense of increasing the
kinetic energy.  
This increase can be understood as a consequence of the uncertainty
principle; the increased correlation allows the charge density to
contract towards the nucleus, which lowers $V_{\rm en}$.  However, the
more localized charge density leads to an increase in the kinetic
energy.  Such a mechanism is not well described by the DCI (doubly
excited CI) wave functions employed in Pauncz's calculations because
the single excitation terms which are excluded contribute to the
modification of the charge density.  The spurious decrease in the
kinetic energy seen in Pauncz's results might be attributed to this.
In the QMC approach, on the other hand, such a modification of the
charge density can be described by the one-body term in the Jastrow
function \cite{see_ref_foulkes} even within single determinant
calculations.

Comparing the DMC and extrapolated DMC results for the singlet and
triplet states we find
\begin{eqnarray}
V_{\rm {ee}}^{S=1} & > & V_{\rm {ee}}^{S=0} \nonumber \\
V_{\rm {en}}^{S=1} & < & V_{\rm {en}}^{S=0} \nonumber \\
T^{S=1} & > & T^{S=0}.
\end{eqnarray}
Both the electron-electron repulsion and kinetic energy are lower in
the singlet state, and the stabilization of the triplet occurs via the
larger reduction in the electron-nucleus energy, corresponding to a
contraction of the electron density towards the nucleus.  This
contraction is promoted by the presence of more spin-parallel pairs in
the triplet which enables the contraction to occur without a large
increase in the electron-electron interaction energy.  The higher
values of $V_{\rm {ee}}$ and $T$ in the triplet state are a
consequence of the contraction depending on the spin state, which
cannot be described within Slater's frozen orbital model.

Our results, which include a high-level description of correlation and
satisfy the virial theorem to high accuracy, support Boyd's less
screening interpretation of Hund's multiplicity rule for the carbon
atom.  It would not be difficult to extend this study to heavier atoms
as, for example, all-electron DMC calculations of Xe ($Z$=54) have
recently been performed.\cite{ma_2004}

\section*{ACKNOWLEDGMENTS}

Our calculations were performed using the facilities of the Center for
Computational Materials Science (IMR; Institute for Materials
Research, Tohoku University), the Center for Information Science of
JAIST (Japan Advanced Institute of Science and Technology), Kansai
Research Establishment (JAERI; Japan Atomic Energy Research Institute)
and the Numerical Materials Simulator of the Computational Materials
Science Center (NIMS; National Institute for Materials Science).  The
authors would like to thank Professor Teruo Matsuzawa (JAIST) for the
generous provision of computational facilities.  K.H and R.M. would
like to thank Dr. Taizo Sasaki for useful conversations.

\newpage

\begin{table}
\caption{Energies for the triplet and singlet states of the carbon
atom.  All energies are given in hartrees. Standard errors in the QMC
results are indicated in parentheses.}
\label{table:2}
\begin{tabular}{cccccccc} \hline
Method & State & Energy & $V_{\mathrm {ee}}$ & $V_{\mathrm {en}}$ & $T$ & Virial ratio \\ \hline
\hline
$\mathrm {HF^a}$ & Triplet & -37.689 & 12.760 & -88.137 & 37.689 & 2.000 \\
 & Singlet & -37.631 & 12.728 & -87.992 & 37.632 & 2.000 \\ 
$\mathrm {CI^b}$ & Triplet & -37.760 & 12.563 & -87.956 & 37.635 & 2.003 \\
 & Singlet & -37.703 & 12.535 & -87.821 & 37.583 & 2.003 \\
VMC & Triplet & -37.7717(1) & 12.629(1) & -88.174(13) & 37.773(12) & 2.000(1) \\
 & Singlet & -37.6896(1) & 12.553(1) & -87.932(13) & 37.689(13) & 2.000(1) \\
DMC & Triplet & -37.8267(4) & 12.583(3) & -88.203(13) & 37.793(12) & 2.001(1) \\
(mixed estimator) & Singlet & -37.7623(6) & 12.509(5) & -87.997(22) & 37.726(19) & 2.001(1) \\
DMC & Triplet &  & 12.538(6) & -88.231(29) & 37.811(26) & 2.002(2) \\
(linear estimator) & Singlet &  & 12.464(9) & -88.062(45) & 37.763(41) & 2.002(2) \\ 
\hline
$^{\mathrm a}$Reference \cite{Katriel2}
$^{\mathrm b}$Reference \cite{Pauncz}
\end{tabular}
\end{table}

\end{document}